# Graphene Composites as Efficient Electromagnetic Absorbers in the Extremely High Frequency Band

Zahra Barani[1,2], Fariborz Kargar[1,2], Konrad Godziszewski[3], Adil Rehman[4], Yevhen Yashchyshyn[3,4], Sergey Rumyantsev[4], Grzegorz Cywiński[4,5], Wojciech Knap[4,5], and Alexander A. Balandin[1,2,§]

[1]Nano-Device Laboratory (NDL), Department of Electrical and Computer Engineering, University of California, Riverside, California 92521 USA

[2]Phonon Optimized Engineered Materials (POEM) Center, Materials Science and Engineering Program, University of California, Riverside, California 92521 USA

[3]Institute of Radioelectronics and Multimedia Technology, Warsaw University of Technology, Warsaw 00-665 Poland

[4]CENTERA Laboratories, Institute of High-Pressure Physics, Polish Academy of Sciences, Warsaw 01-142 Poland

[5]CEZAMAT, Warsaw University of Technology, 02-822, Warsaw, Poland

---

[§] Corresponding author (A.A.B.): balandin@ece.ucr.edu ; web-site: http://balandingroup.ucr.edu/

**Abstract**

We report on the synthesis of the epoxy-based composites with graphene fillers and testing their electromagnetic shielding efficiency by the quasi-optic free-space method in the extremely high frequency (EHF) band (220 – 325 GHz). The curing adhesive composites were produced by a scalable technique with a mixture of single-layer and few-layer graphene layers of a few-micron lateral dimensions. It was found that the electromagnetic transmission, $T$, is low even at small concentrations of graphene fillers: $T < 1$ % at frequency of 300 GHz for a composite with only $\phi = 1$ wt% of graphene. The main shielding mechanism in composites with the low graphene loading is *absorption*. The composites of 1 mm thickness and graphene loading of 8 wt% provide excellent electromagnetic shielding of 70 dB in the sub-terahertz EHF frequency with negligible energy reflection to the environment. The developed lightweight adhesive composites with graphene fillers can be used as electromagnetic absorbers in the high-frequency microwave radio relays, microwave remote sensors, millimeter wave scanners, and wireless local area networks.

**I. Introduction**

Rapid development of the wireless communications, distributed sensor arrays, and portable electronic devices made the control of electromagnetic (EM) radiation and reduction of EM pollution crucially important.[1–3] The electromagnetic interference (EMI) shielding is needed to ensure that electronics operate reliably and without detrimental effects on human health.[4–8] Different frequency ranges and types of devices require different solutions for EMI shielding. The shielding of EM energy at high frequency bands can be particularly challenging. In many cases, additional requirements are imposed on the materials used for EMI shielding, including thickness and weight limits, mechanical and thermal properties, electrical conduction or isolation. Absorption of EM energy rather than its reflection back to the environment offers benefits in a wide range of commercial and defense applications. However, many existing EMI shielding materials, *e.g*. metallic coatings, simply redirect the EM energy via electrical conduction – based reflection mechanism. The latter shifts the problem of EM pollution from one element to another, and while protecting electronic components, it can negatively affect the human health.[4–8] The metal-based EMI shielding materials have other problems associated with heavy weight, corrosion, and difficulty of processing.

An alternative approach to metal shielding is the use of polymer-based materials with electrically conductive fillers.[9–16] Such composite materials demonstrated effective EMI shielding in MHz and lower GHz ranges. The initial research was conducted on composites with metal particles, which were added as fillers in high weight fractions in order to increase the electrical conductivity.[10,13,16–19] However, the polymer-metal composites suffer from relatively high weight, cost and corrosion. More recent studies reported the use of carbon fibers,[20–28] carbon black,[28,29] bulk graphite,[30–32] carbon nanotubes,[33–39] reduced graphene oxide,[40–52] graphene[53–56] and, combination of carbon allotropes with or without other metallic –or non-metallic particles.[42,44,53,55,57–62] A new class of quasi-two-dimensional materials, MXene's, have also been shown to exhibit high EM shielding efficiencies when added as fillers to polymer matrices or deployed as thin membranes.[17] However, the dominant mechanism of EMI shielding in these materials is reflection, which make them less attractive for a wide range of applications where EM reflection creates problems. The studies of graphene-enhanced composites for EMI shielding mostly focused on super high frequency (SHF) range, specifically X-band.[51–56,63,64] The reported studies covered the range of frequencies from MHz to 30 GHz. We are aware of one study of the use of graphene composites in sub-THz range.[65] One can conclude from the reported studies that effectiveness and physical mechanisms of shielding differ substantially depending on the EM frequency range, graphene loading fraction, and characteristics size and thickness of the fillers. In our prior work, we demonstrated a dual function of graphene composites for the X-band, which include EMI shielding and heat removal.[3] We are not aware of the reports, which would address the EM wave interaction with the polymer-based graphene composites in the sub-terahertz extremely high frequency (EHF) band.

Here, we report the results of investigation of the EMI shielding efficiency of graphene epoxy-based composites in the EHF band with the frequencies from 220 GHz to 320 GHz, which correspond to the WR-3 band in microwave wave guide classification. This sub-terahertz frequency band is important for radio astronomy, high-frequency microwave radio relay, microwave remote sensing, millimeter wave scanning, and wireless local area networks. From the physics point of view, this frequency band is interesting because of possible changes in the electrical percolation in composites as compared to percolation in DC or low-frequency regime. We found that the synthesized graphene-enhanced composites are not only efficient EMI shielding materials but also achieve their function by *absorbing* EMI waves rather than by *reflecting* them. Moreover, it appears that there exists an optimum loading fraction of graphene, near 1 wt %, at which the material effectively absorbs but not reflects the EM waves. The examined epoxy-based curing composites can serve as adhesives for the electronic components while performing the EMI shielding functions.

## II. Material Synthesis

In the context of present study, we use the term *graphene* for a mixture of single-layer graphene (SLG) and few-layer graphene (FLG). The typical lateral dimensions of the SLG and FLG flakes are in a few-micrometer range. More detailed analysis of the size and thickness distribution of SLG and FLG in a mixture has been reported by some of us earlier in the context of thermal studies.[66–70] For this work, we utilized commercially available graphene (xGnP®H-25, XG-

Sciences, U.S.A.) to prepare the composites. An in-house designed mixer was used to disperse the graphene fillers uniformly in the high loading composites.[71] The samples were prepared in the form of disks with the diameter of 25.6 mm and thicknesses from 0.9 mm to 1.0 mm. The sample thickness affects the total absorption and the total shielding efficiency of the composites. The schematic of the shielding function of the materials, optical images of the samples, and Raman spectroscopy data are presented in Figure 1. The Raman spectrum of representative composite samples confirms the composition *via* the characteristic peaks of the epoxy and graphene, as well as the expected change in concentration evidenced from the intensity increase of the G peak and 2D band of graphene in the samples with the higher graphene loading. Additional material characterization data for the graphene-enhanced composites, including representative scanning electron microscopy (SEM) images are provided in the Supplemental Materials.

[Figure 1]

### III. Results of the EM Measurements

The shielding efficiency of the epoxy-based graphene composites was determined from the scattering parameters measured using the quasi-optic free space method.[72–74] The measurement setup consisted of a vector network analyzer with a pair of frequency extenders – two high gain horn antennas and two double convex lenses. The measurements were performed in the range of frequencies from 220 GHz to 320 GHz. In order to obtain the reflection, *R*, and transmission, *T*, coefficients three measurements were performed. The first measurement was conducted with the sample, the second one was without the sample, and the third measurement was with a reference plane metal reflector. Two last measurements were used as the references to calculate the transmission and reflection coefficients, respectively. The reference measurements allow one to compensate for the transmission losses in the measurement path. The transmission and reflection coefficients are calculated according to the standard equations[37]

$$T = \frac{|S_{21s}|^2}{|S_{21e}|^2}, \tag{1}$$

$$R = \frac{|S_{11s}|^2}{|S_{11m}|^2}. \tag{2}$$

Here $S_{11s}$ and $S_{21s}$ are the results for the measurements with the sample, $S_{21e}$ is the result for the measurement with an empty optical path, $S_{11m}$ is the result for the measurement with a metal plate.

In the case of multiple reflections in the quasi-optical path, the measurement data can be affected. To account for this possibility, an additional data processing step was applied. It consisted of the time domain gating.[75] The latter was possible owing to the broad frequency range and a large number of the measurement points (up to 32000). The measured complex scattering parameters were transformed to the time domain. After that an appropriate time domain window was applied. Finally, the time gated data were transformed back to the frequency domain. This approach allowed us to improve the accuracy and reliability of the obtained data for the transmission and reflection coefficients. The obtained $R$ and $T$ coefficients were used for calculation of the absorption coefficient, $A$, and the effective absorption coefficient, $A_{eff}$, which are given as[37]

$$A = 1 - R - T, \tag{3}$$

$$A_{eff} = \frac{A}{1-R} = \frac{1-R-T}{1-R} = \frac{A}{A+T}. \tag{4}$$

Both parameters define the EM absorption characteristics of the shielding material. One should note that $A_{eff}$ describes the actual absorption properties of the material since part of the EM energy is reflected from the surface of the material.

Figures 2 (a) and (b) show the reflection, $R$, and transmission, $T$, coefficients for pristine epoxy and epoxy-based composites with the graphene loading ranging from 0.8 wt % to 8.0 wt%. Figure 2 (c) shows the lower bound of the transmission coefficient ($T < 5\%$) specifically allowing to distinguish the transmission coefficients for the composites with the graphene loading above 1 wt %. One can see that the a small addition of graphene (0.8 wt % − 1.0 wt %) does not change noticeably the EM wave reflection from the samples but does significantly reduce the transmission. The transmission coefficient decreases from 60 % for the pristine epoxy to less than 5 % for the epoxy-based composite with only 1 wt % of graphene fillers (see both Figure 2 (b) and (c)). The transmission monotonically decreases with the frequency for the composites with the graphene loading of 1 wt % or more in the considered frequency band.

[Figure 2]

Figures 3 (a) and (b) show the absorption coefficient, $A$, and the effective absorption coefficient, $A_{eff}$, respectively. Both parameters define the EM absorption characteristics of the shielding material. Figure (c) shows the upper bound of the effective absorption coefficient ($A_{eff} > 95\%$), specifically allowing to distinguish the values of $A_{eff}$ for the composites with the graphene loading above 1 wt %. The data in Figure 3 make it clear that the main EM wave shielding mechanism in composites with the low graphene loading (0.8 wt % − 1.0 wt %) is absorption. The absorption coefficient, $A$, increases from 20 % for the pristine epoxy to more than 80 % for the epoxy-based composite with only 1 wt % of graphene fillers (see Figure 3 (a)). The absorption monotonically increases with the frequency for the composites with the graphene loading of 1 wt % or more in this frequency band (see both Figure 3(b) and 3(c)). The data presented in Figures 2 and 3 indicate that composites with the low loading of graphene provide efficient EMI shielding in this frequency band *via* absorption with small energy reflection to the environment. The graphene loading of 1 wt % appears to be optimum for the tested composites with the given size and thickness of the fillers.

[Figure 3]

Figure 4 (a-b) presents the shielding efficiency of the composites by reflection, $SE_R$, and absorption, $SE_A$, respectively. The latter includes the internal reflections of the EM waves inside the composite medium. The total shielding efficiency, $SE_T$, is the sum of shielding by reflection and absorption, and it is plotted as a function frequency for different filler loadings in Figure 4 (c). For these plots, the shielding efficiency parameters were calculated from the measured $R$ and $A_{eff}$ using the following equations[3]

$$SE_R = -10 \log(1 - R), \quad (5)$$
$$SE_A = 10 \log(1 - A_{eff}), \quad (6)$$
$$SE_T = SE_R + SE_A. \quad (7)$$

[Figure 4]

As one can see in Figure 4 (a), with addition of graphene fillers up to $\phi \leq 1$ wt%, $SE_R$ does not change, and then gradually increases from ~0.6 dB to ~3 dB at $\phi = 8$ wt%. In contrast, $SE_A$ increases continuously with graphene loading and reaches ~40 dB at $\phi = 8$ wt% at the frequency of $f = 220$ GHz (see Figure 4 (b)). Note that $SE_A$ is constant for the pristine epoxy in entire frequency range. In contrast, in composites with even a small loading of graphene, $SE_A$ increases as a function of frequency. The higher the graphene loading into the matrix, the higher is the rate of the increase. For the composite with $\phi = 8$ wt%, $SE_A$ approaches 70 dB at $f = 320$ GHz, which is beyond the requirements of most of the industrial applications. The classical Simon's equation relates $SE_A$ [dB] to frequency, thickness, and electrical resistivity as follows[3,9,17]

$$SE_A = (1.7t/\rho) f^\gamma, \quad (8)$$

where $t$ [cm] is the thickness, $\rho$ [$\Omega$ cm] is the bulk resistivity, and $f$ [MHz] is the frequency, respectively. Simon's equation assumes a constant frequency exponent $\gamma = 0.5$. Our attempts to fit the experimental data, presented in Figure 4 (b), with Eq. (8) and the fixed value of $\gamma = 0.5$ were not successful due to a strong dependence of $SE_A$ on frequency in composites with graphene. We changed the procedure by treating both $\rho$ and $\gamma$ as the fitting parameters. The latter resulted in obtaining a nearly constant value of $\gamma = 1.44$ for all concentrations and different values of the effective electrical resistivity for different filler loadings $\phi$. The agreement of the fittings with the experimental data was excellent (see Supplemental Materials). For the composites with graphene, the above equation in the frequency range between 220 GHz $\leq f \leq$ 320 GHz can be revised as

$$SE_A = (1.7t/\rho)\,f^{1.44}. \tag{9}$$

The extracted values of $\rho$ and $\gamma$ are listed in Table I. We interpret the extracted $\rho$ as an effective high-frequency resistivity. As described below, the obtained values are reasonable and consistent with DC resistivity measurements.

| Table 1: Resistivity and frequency exponent for epoxy with graphene fillers | | | |
|---|---|---|---|
| $\phi$ [wt%] | $t$ [cm] | $\rho$ [$\Omega$cm] | $\gamma$ |
| 0.8 | 0.1 | $3.61 \times 10^6$ | 1.495 |
| 1.0 | 0.1 | $5.65 \times 10^5$ | 1.426 |
| 4.0 | 0.1 | $2.95 \times 10^5$ | 1.417 |
| 8.0 | 0.1 | $1.51 \times 10^5$ | 1.415 |

In Figure 5 we present the absorption shielding efficiency, $SE_A$, as the function of the graphene filler loading near the optimum loading of ~1 wt % for the given fillers. The data are presented at two frequencies of 250 GHz and 300 GHz. One can see a strong increase in the shielding efficiency at this loading. At higher graphene loading, the shielding efficiency increases slower while the reflection grows rapidly. This trend explains the existence of the optimum graphene concentration in the composites for EMI absorbers in this frequency band. We note that there was a substantial $SE_A$ data scatter at the transition point of ~1 wt %. For this reason, the data points in this region were averaged over results obtained in several measurement runs.

[Figure 5]

We have measured the DC resistance of the graphene-enhanced composites in the vertical and in-plane, *i.e.* lateral directions. For the vertical resistance measurements two high area contacts made of conducting paste were deposited on top and bottom parts of the sample. The resistance was measured in the two contacts configuration. The large cross-sectional area of the contacts allowed us to measure high resistivity of up to ~$10^{13}$ [$\Omega$cm]. The in-plane resistance was measured in the 4-probe configuration. For this purpose, four needles contacted the surface of the sample using the probe station. The distance between the contacts was ~0.5 mm. The resistivity was calculated, in a standard way, as $\rho = 2\pi s F R$, where $s$ is the distance between contacts, $F = 0.8$ is the correction factor to take into account the finite thickness of the sample, and $R$ is the measured resistance.[75] Figure 6 shows the resistivity of the samples as a function of the graphene filler loading. The shaded region indicates the region, pertinent to the low loading samples, where the resistance was too high to measure in the 4-probe configuration. The in-plane resistivity data are shown for the center and peripheral regions of the samples. The in-plane resistivity in the center was consistently slightly smaller than at the edges. One should note that the in-plane resistivity is always smaller than the vertical one. This is an indication that graphene fillers are not completely oriented in random directions and have some preference for the in-plane orientation. This conclusion is in line with the SEM inspection of the representative samples (see Supplementary Materials). The preferential in-plane orientation can provide positive effect on the EMI shielding properties of the composites. Some degree of the graphene filler alignment was achieved naturally, as a result of the composite preparation procedure, without extra technological steps (see the Methods). The measured DC resistivity for composites with graphene are of the same order of magnitude as those obtained with Eq. (9). The latter provides an additional validation to our fitting procedure and the formula. The DC resistivity is not expected to be exactly the same as the high-frequency effective resistivity.

The important observation is that the resistivity of composites with the graphene loading below 5 wt % is high. The material is electrically insulating, which is beneficial for many EMI shielding applications. The electrical conductivity of the composite is not required for the EM energy absorption and reflection because EM wave can couple locally to the graphene fillers. We also note that the DC data in Figure 6 does not show an abrupt onset of electrical percolation. There is a rather graduate decrease in the electrical resistivity with increasing graphene loading. From the other side, in Figure 5, we see a drastic increase in $SE_A$ at the loading of 1 wt% - 2 wt%. We hypothesis that the increase in $SE_A$ is related to on-set of electrical percolation at extremely high frequency band. The physics of percolation at high frequency is expected to be different than that at DC or low frequency because of possible contributions of the displacement currents between the graphene fillers separated by narrow dielectric layers. The on-set of the electrical percolation at high frequency will lead to the enhanced free-electron type reflection for the whole composite sample. However, more studies are needed to make a definitive conclusion about the mechanism of electrical percolation at high frequency in this type of composites.

[Figure 6]

The epoxy-based curing composites can serve as adhesives for the electronic components while simultaneously performing the EMI shielding functions. In this work, we focused on composites with the low graphene loading in order to provide EMI shielding without EM wave reflection to environment. At this graphene concentrations, the thermal conductivity of the composites is not substantially different from those of the pristine epoxy.[66] The increase of graphene loading results in strong enhancement of the heat conduction properties of epoxy-based composites.[66] The latter results from excellent thermal conductivity of graphene and FLG.[76–79] From the other side, higher loading of graphene fillers also increases the reflection of EM waves as the data in Figure 2 and 3 show. The strong absorption of EM wave by the composites means that the energy is transferred from EM waves to heat. In this sense, the higher thermal conductivity of composites can be beneficial. One can envision possible approaches for increasing the thermal conductivity without increasing the EM wave reflection. The first one is the use of layered composites where the inside layer has higher graphene loading for better heat conduction while the outside layer has graphene loading of ~1 wt % for optimum EMI shielding without reflection. The second approach is to use a combination of graphene and boron nitride (BN) fillers with larger total loading. The BN fillers, unlike graphene fillers, are not electrically conductive but also have much higher thermal conductivity than epoxy base.[66,67]

## V. Conclusions

We reported on the synthesis of the epoxy-based composites with graphene fillers and testing their EMI shielding efficiency in the sub-terahertz frequency range (band: 220– 320 GHz). It was found that the electromagnetic transmission, T, is low even at small concentrations of graphene fillers: T < 1 % at frequency of 300 GHz for a composite with 1 wt % of graphene fillers. Our results demonstrate that graphene epoxy-based composites provide efficient electromagnetic shielding in the sub-terahertz band *via* absorption and small energy reflection to the environment. At the frequency 320 GHz, we achieved the total shielding efficiency of ∼70 dB for epoxy with only 8 wt% graphene loading. This EMI shielding is beyond the requirements of most of the industrial applications. The developed lightweight composites with graphene can be used as electromagnetic absorbers in the microwave radio relays, remote sensors, millimeter wave scanners, and wireless local area networks. Since our composites are curing materials, they can be used as adhesives for packaging the electronic components while simultaneously performing the EMI shielding functions.

**METHODS**

**Sample Preparation:** The composite samples were prepared by mixing the commercially available FLG flakes (xGnP®H-25, XG-Sciences, US) with epoxy (Allied High Tech Products, Inc.) using a high-shear speed mixer (Flacktek Inc.) at 800 rpm and 2000 rpm each for 5 minutes. The mixture was vacuumed for 30 minutes. After that time, the curing agent (Allied High Tech Products, Inc.) was added in the mass ratio of 12: 100 with respect to the epoxy resin. The compound was mixed and vacuumed one more time and left in the oven for ~2 hours at 70 °C in order to cure and solidify.

*Acknowledgements*

The work at UC Riverside was supported, in part, by the by the Office of Technology Partnerships (OTP), University of California via the Proof of Concept (POC) project "Graphene Thermal Interface Materials" and by the UC - National Laboratory Collaborative Research and Training Program - University of California Research Initiatives LFR-17-477237. This work was also supported by CENTERA Laboratories in frame the International Research Agendas program for the Foundation for Polish Sciences co-financed by the European Union under the European Regional Development Fund (No. MAB/2018/9) and partially supported by the Foundation for Polish Science through the TEAM project POIR.04.04.00-00-3D76/16 (TEAM/2016-3/25).

**Contributions**

A.A.B. and S.R. conceived the idea of the study, coordinated the project, and contributed to the EM data analysis. K.G. and Y.Y. performed the EM shielding measurements. Z.B. prepared the composites, performed materials characterization and assisted with the EM data analysis; F.K. contributed to the sample preparation and EM data analysis. A.R. performed DC resistivity measurements. G.C. and K.W. contributed to data analysis. A.A.B. and S.R. led the manuscript preparation. All authors contributed to writing and editing of the manuscript.

Barani *et al*., Graphene Composites as Efficient Electromagnetic Absorbers in the Extremely High Frequency Band (2020)

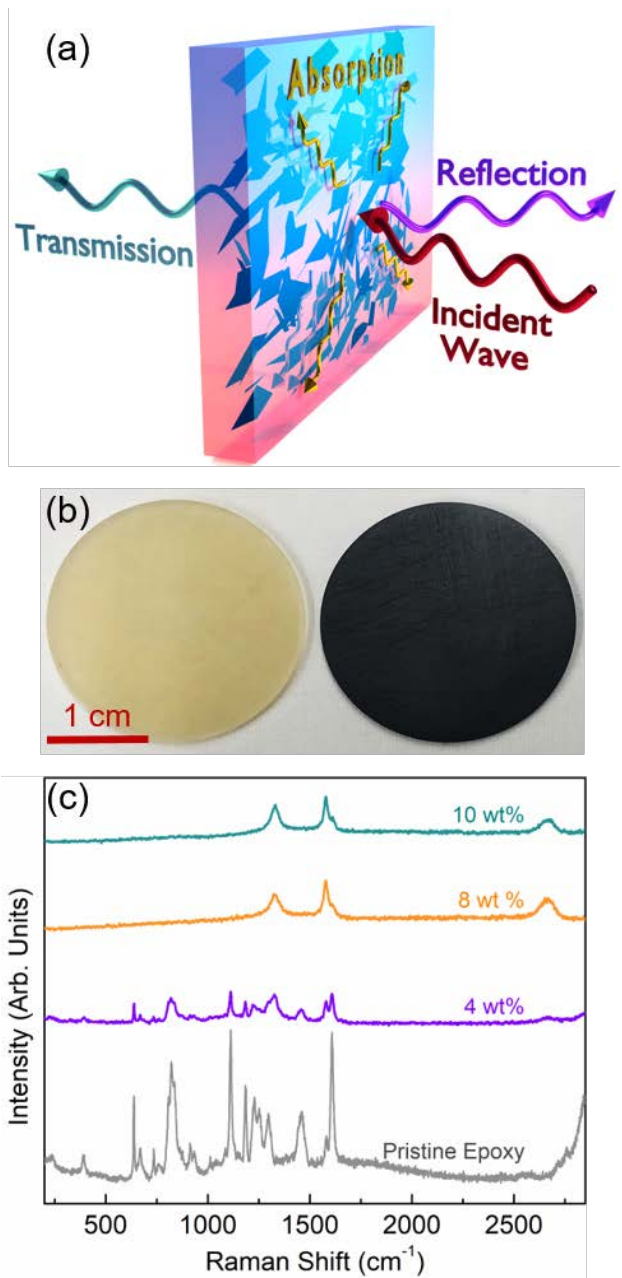

**Figure 1:** (a) Schematic of the interaction of the EM waves with the epoxy-based composites with the low filler loading of graphene. (b) Optical image of the pristine epoxy (left) and epoxy with $f = 4$ wt% of graphene (right). (c) Raman spectra of epoxy with different loading fractions of graphene fillers. Note the color change of the epoxy composite with addition of graphene. The composites can be used as adhesive for mouting microwave devices and as a coating layer.

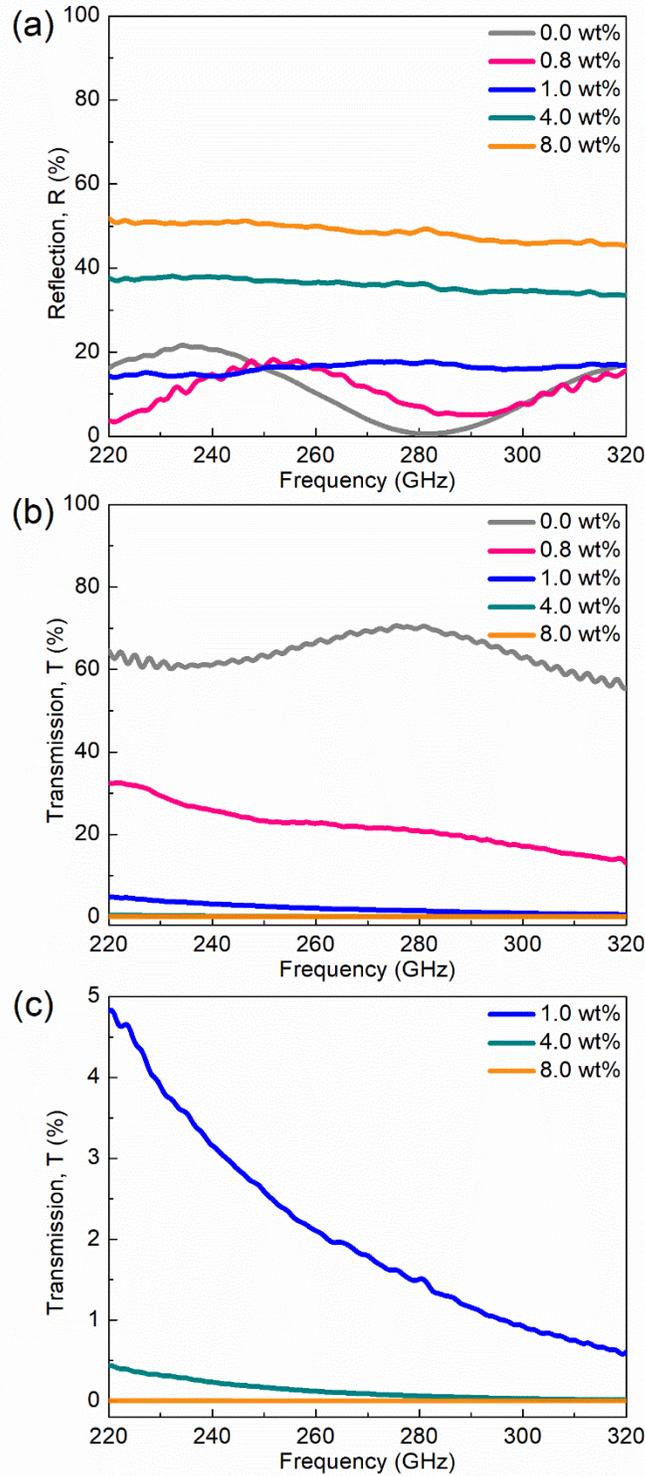

**Figure 2:** Coefficients of (a) reflection and, (b) transmission for composites with different graphene loadings in the 220 GHz – 320 GHz frequency range. (c) The lower bound of the transmission coefficient ($T < 5\%$) allowing to distinguish the transmission coefficients for the composites with the graphene loading above 1 wt. %.

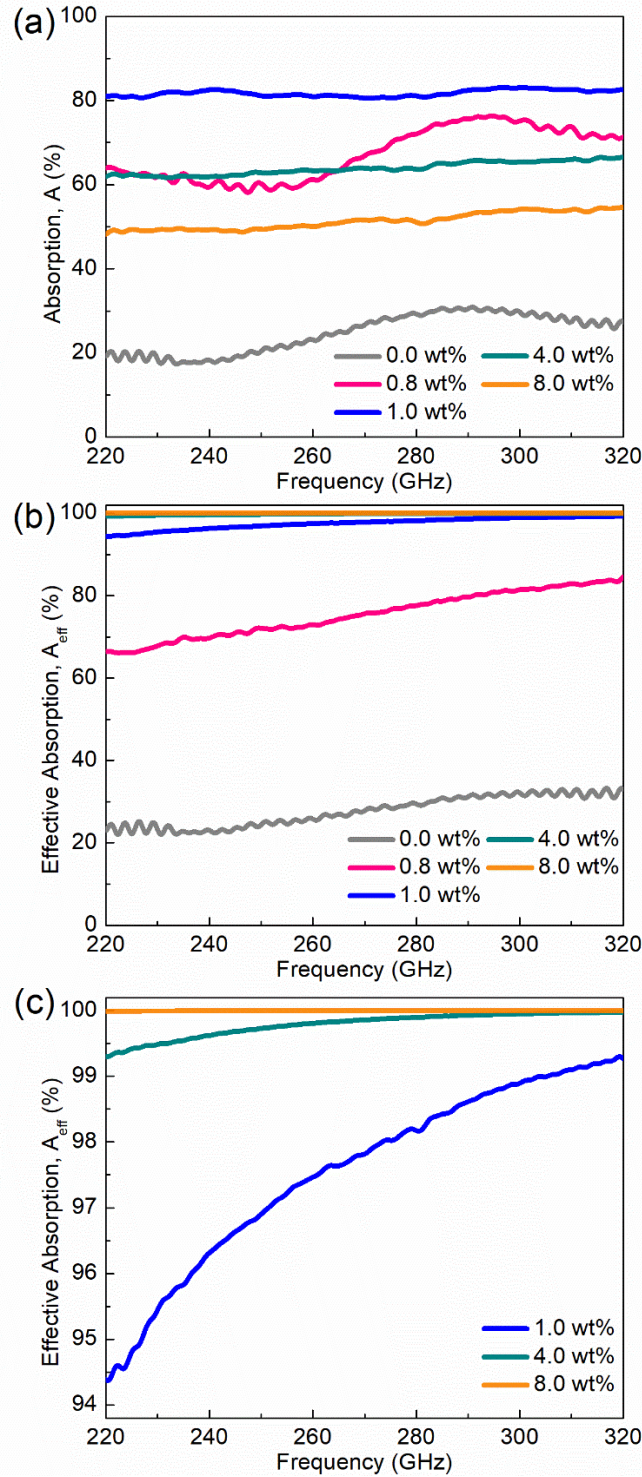

**Figure 3:** Coefficients of (a) absorption and (b) effective absorption for composites with different graphene loading fractions. (c) The upper bound of the effective absorption coefficient ($A_{eff} > 95\%$) allowing to distinguish the values of $A_{eff}$ for the composites with the graphene loading above 1 wt. %.

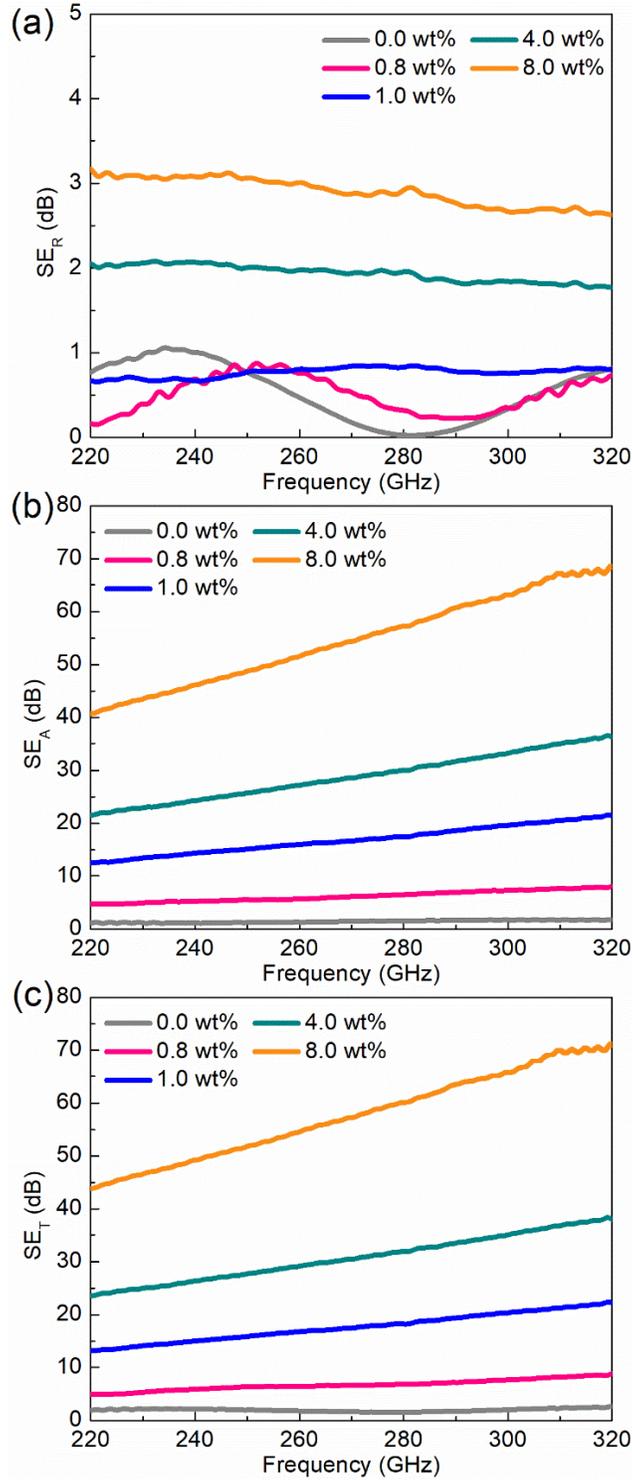

**Figure 4:** (a) Reflection ($SE_R$), (b) absorption ($SE_A$), and (c) total ($SE_T$) shielding efficiencies of composites at different graphene concentrations. As the filler loading increases, $SE_R$ does not grow significantly wheras $SE_A$ increasese substantially. The total shielding efficiency is substancially increased as a result of $SE_A$ enhancement.

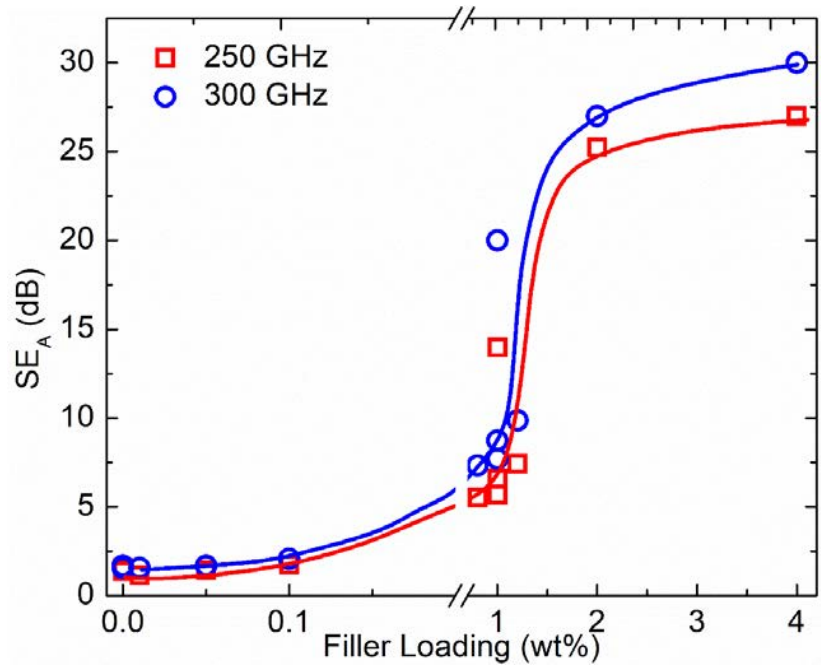

**Figure 5:** Absorption shielding efficiency as a function of graphene filler loading. Note an abrupt increase in $SE_A$ at the filler loading, $f$, in the interval wt% $< f <$ 2 wt%.

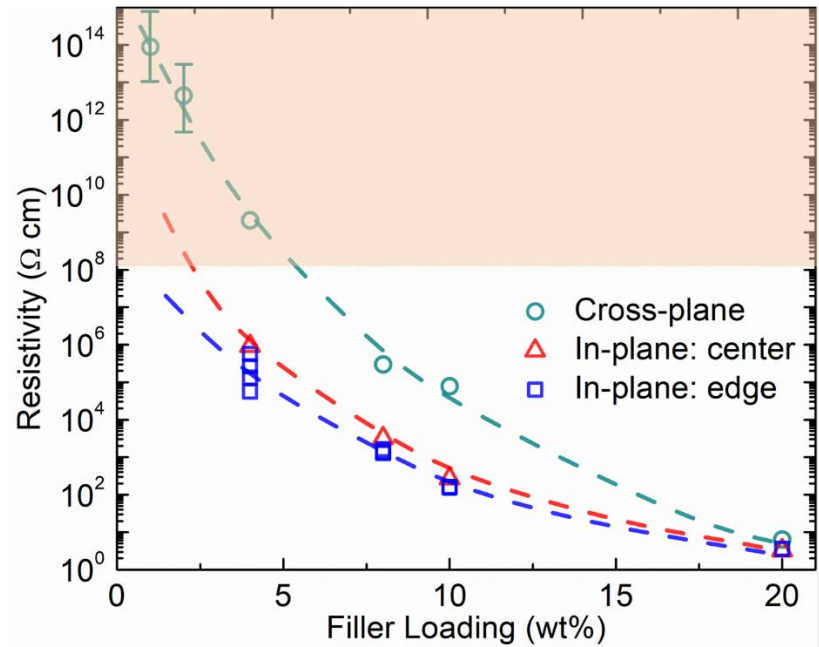

**Figure 6:** In-plane and cross-plane electrical resistivity of the samples as a function of graphene loading fraction measured at DC bias. The in-plane resistivity is smaller than that of the cross-plane direction, indicative that the fillers have preferential in-plane orientation as a result of the sample preparation procedures.